**Superoxide anion ($O_2^-$) collisions with $CO_2$ molecules in the energy range 50-950 eV**

C. Guerra[1], M. Leiferman[2], A.I. Lozano[1,3,4], F. Aguilar-Galindo[2,5], S. Díaz-Tendero[2,5,6], J. C. Oller[7], P. Limão-Vieira[3] and G. García[1,8]

[1]Instituto de Física Fundamental, Consejo Superior de Investigaciones Científicas, Serrano 113-bis, 28006 Madrid, Spain.
[2]Departamento de Química, Módulo 13, Universidad Autónoma de Madrid, 28049 Madrid, Spain.
[3]Laboratório de Colisões Atómicas e Moleculares, CEFITEC, Departamento de Física, Universidade NOVA de Lisboa, 2829-516 Caparica, Portugal.
[4]Institut de Recherche en Astrophysique et Planétologie (IRAP), Université Toulouse III - Paul Sabatier, 9 Avenue du Colonel Roche, Toulouse 31028, France.
[5]Institute for Advanced Research in Chemistry (IAdChem), Universidad Autónoma de Madrid, 28049 Madrid, Spain.
[6]Condensed Matter Physics Center (IFIMAC), Universidad Autónoma de Madrid, 28049 Madrid, Spain.
[7]División de Tecnología e Investigación Científica, Centro de Investigaciones Energéticas, Medioambientales y Tecnológicas, 28040 Madrid, Spain.
[8]Centre for Medical Radiation Physics, University of Wollongong, NSW 2522, Australia

**Abstract**

A novel gas-phase molecular scattering study is reported for $O_2^-$ colliding with $CO_2$ for impact energies ranging from 50 to 950 eV. The absolute total electron detachment, relative total and partial ionization cross sections have been measured within this energy range and the positive ion yield of those produced during the collisions has been obtained. The primary anionic beam projectile is produced in a pulsed hollow cathode discharge induced plasma, and its interactions with the neutral molecular target occur in a gas cell at a well-known constant pressure. For impact energies above 500 eV high mass (m> 44 u) charged complexes have been detected. With the aid of a theoretical study, using ab initio methods, we propose a mechanism to infer on the formation of these cationic species, which have been assigned as projectile-target stable compounds ($CO_3^+$ and $CO_4^+$).

**1. Introduction**

The $CO_2$ molecule has acquired increasing interest in recent decades due to its relevance in laser technology [1] with significant applications in medicine [2]. It plays an important role in plasma diagnostics [3-4], solving atmospheric [5] and astrophysical issues [6] as well as modelling particle transport in gas mixtures [7]. The atmospheric concentration of $CO_2$ affects basically to plant metabolism [8,9], global warming [10,11] and other environmental parameters directly related to the stability of the biosphere [12].
$CO_2$ electronic structure and dissociation patterns have been studied through photoabsorption experiments [13-15]. Electron and positron scattering by $CO_2$ have also been the subject of many theoretical and experimental studies and the interaction cross section data have been recently summarized [16].

Negative ions ($H^-$, $D^-$, $He^-$, $F^-$, $Cl^-$, $I^-$, $NO^-$, and $O^-$) colliding with $CO_2$, have been studied by different authors, paying special attention to electron detachment, charge exchange and resonant processes [17-21]. Paulson [22] conducted a comprehensive study on the reactions $O^- + CO_2 \rightarrow CO_2^- + O$, $O^- + CO_2 \rightarrow O_2^- + CO$, $NO^- + CO_2 \rightarrow CO_2^- + NO$, $O_2^- + CO_2 \rightarrow CO_3^- + O$, and $O_2^- + CO_2 \rightarrow CO_2^- + O_2$, thus obtaining effective cross section values for energy ranges between 1 and 100 eV. Negative ion reaction with different molecular targets, including $CO_2$ were also reported by Moruzzi and Phelps [23]. Doverspike et al. [24] reported absolute electron detachment cross sections for colliding anionic projectiles ($Cl^-$ and $Br^-$) with different neutral molecular targets $H_2$, $D_2$, $O_2$, $N_2$, $CO$, $CO_2$, and $CH_4$, in the energy range between 1 and 300 eV.

As far as cation collisions with $CO_2$ studies are concerned, we note that have been extensively discussed in the literature. Greenwood et al. [25] implemented a novel experimental beam line to study a certain type of collisions that occur in the interaction of solar wind ions with cometary gases, obtaining absolute measurements of the charge-exchange cross sections for collisions of highly charged ions ($H^+$, $He^+$ and $He^{2+}$) with neutral targets ($H_2O$ and $CO_2$) in the impact energy range 0.3-7.5 keV.

The superoxide anion ($O_2^-$) is one of the reactive oxygen species (ROS) whose significance in diverse chemical and biological systems attracted the interest of many investigations [26]. Due to their reactive properties, they play an important role in biomedical [27, 28] and industrial [29] applications. The chemistry of $O_2^-$ in space has also been recently analyzed [30]. However, superoxide anion scattering cross section data for relevant biological and environmental molecules are scarce. In particular, as far as we know, apart from the aforementioned anion reaction studies [22], collisional data of $O_2^-$ with $CO_2$ are not reported in the literature.

In this study we present a theoretical and experimental investigation on collisions of $O_2^-$ with $CO_2$ in the gas-phase. Absolute total electron detachment cross sections (TEDCS) for impact energies ranging from 50 to 950 eV have been measured with a transmission-beam experimental system [31]. We have also obtained relative total ionization cross sections (TICS) by analyzing the ion formation intensities with respect to that of the primary ion beam for each incident anion energy. The induced cationic fragmentation has been calculated by means of quantum chemistry simulations, proposing the mechanism for formation of stable $CO_3^+$ and $CO_4^+$ compounds.

The remainder of the present paper is organized as follows. In section 2, we describe the experimental procedure and analyze the possible uncertainty sources. The theoretical methods are described in Section 3. In section 4, we present and discuss the obtained results. Finally, in Section 5 we summarize the main conclusions of this study.

## 2. Experimental procedure

### 2.1 Apparatus

The experimental system has been entirely described in previous publications [31, 32] and, therefore, we will only briefly present here some operational details.

The incident anion beam is generated in a hollow-cathode plasma discharge produced by a pulsed oxygen molecular beam passing through a 20 mm length cylinder negatively biased (-500V). The neutral oxygen beam is provided by a supersonic expansion valve (Parker VAC1250). Negative ions are produced in the discharge afterglow, are then accelerated by the cathode voltage and focused on the scattering chamber entrance aperture (2 mm in diameter). The scattering chamber is a gas cell (GC) where the molecular target is introduced at low pressure (< 2 mTorr) through a sapphire leak valve. The anionic oxygen beam is there accelerated or decelerated by applying a negative or positive voltage to the GC, thus defining the kinetic energy of the $O_2^-$ anions during the collision. The positive ions produced by the $O_2^-$-$CO_2$ interactions in the GC are extracted and accelerated, by means of an extractive plate system, through a perpendicular time-of-flight (TOF) mass spectrometer (1.40 m length), placed over the collision region. Produced ions are finally detected by a microchannel plate detector (MCP1) operating in single pulse counting mode to facilitate their mass analysis. The extractive parallel plates are connected to positive (0 to +900) and negative (0 to -900) pulsed voltages, respectively. The voltage and duration of these pulses were optimized to ensure a total and uniform ion extraction but maintaining a reasonable mass resolution. In this study, the length of the hollow cathode limits the mass resolution. The uncertainty in the position where anions are formed along the cathode is converted into a time of flight, and therefore mass, uncertainty which determines our inherent mass resolution limit ($\Delta m/m = 0.05$).

The experimental setup also accounts for a homemade electron gun placed under the GC that provides an energy-controlled electron beam (0–500 eV electron incident kinetic energy) entering the interaction region perpendicular to the anion beam and opposite to the TOF mass analyzer. The main purpose of the electron gun is to analyze the molecular composition of the background and the gas target as well as to heat the chamber to facilitate its evacuation when needed. It also provides a reference electron induced fragmentation spectrum to compare with that produced by anion collisions.

Finally, the transmitted anionic beam is perpendicularly repelled by a continuous -250V voltage towards a second microchannel plate detector (MCP2) which monitors the primary anion beam intensity. The MCP2 detector signal is also used to determine the primary beam energy distribution by using a set of three grids as a retarding field analyzer placed at the entrance of the transmission chamber (see ref. 31 and 32 for details).

The projectile anionic beam is mainly formed by $O_2^-$ but also contains certain amounts of $O^-$, and $O_3^-$. Tuning the extractive pulses, we can select the area of the primary beam which is really generating the positive ion fragments being mass analyzed (see ref. 31, 32 and 33 for details). This procedure acts as a TOF filter of the primary anion beam.

## 2.2 Data Acquisition Methodology

### a. Measurement methodology

The TEDCS ($\sigma_t$) for $O_2^-$ collisions with $CO_2$, within the impact energy range (50-950 eV), has been obtained by measuring the primary anion beam attenuation as a function of the target ($CO_2$) pressure from 0 to 2 mTorr. The absolute pressure of the gas target is measured by two capacitance manometers (MKS Baratron 627B) placed at opposite sides of the GC ($P_{GC1}$ and $P_{GC2}$, respectively), perpendicular to the primary beam direction, so ensuring that pressure gradients are not affecting the present measurements. In these

conditions, the TEDCS value is derived from the Beer-Lambert attenuation law assuming an ideal gas behavior:

$$I = I_o \exp\left(-\frac{Pl\sigma_t}{kT}\right),　\quad (1)$$

where $I$ is the transmitted anion intensity through the GC, $P$ is the $CO_2$ pressure $[P=(P_{GC1}+P_{GC2})/2]$, $I_o$ is the initial intensity (no gas in the GC), $l$ is the effective path length of the collision chamber (36 mm), $k$ is the Boltzmann constant, and $T$ is the absolute temperature. The $O_2^-$ transmitted intensity is measured with a single ion counting system (see Ref. 31 for details) as a function of the $CO_2$ gas pressure. For each attenuation measurement, pressure is varied from 0 to a maximum value ($P_{max}$) which is chosen to ensure that we are operating under single collision conditions [31]. Fig. 1 shows some typical attenuation curves for different impact energies ranging from 50 to 900 eV

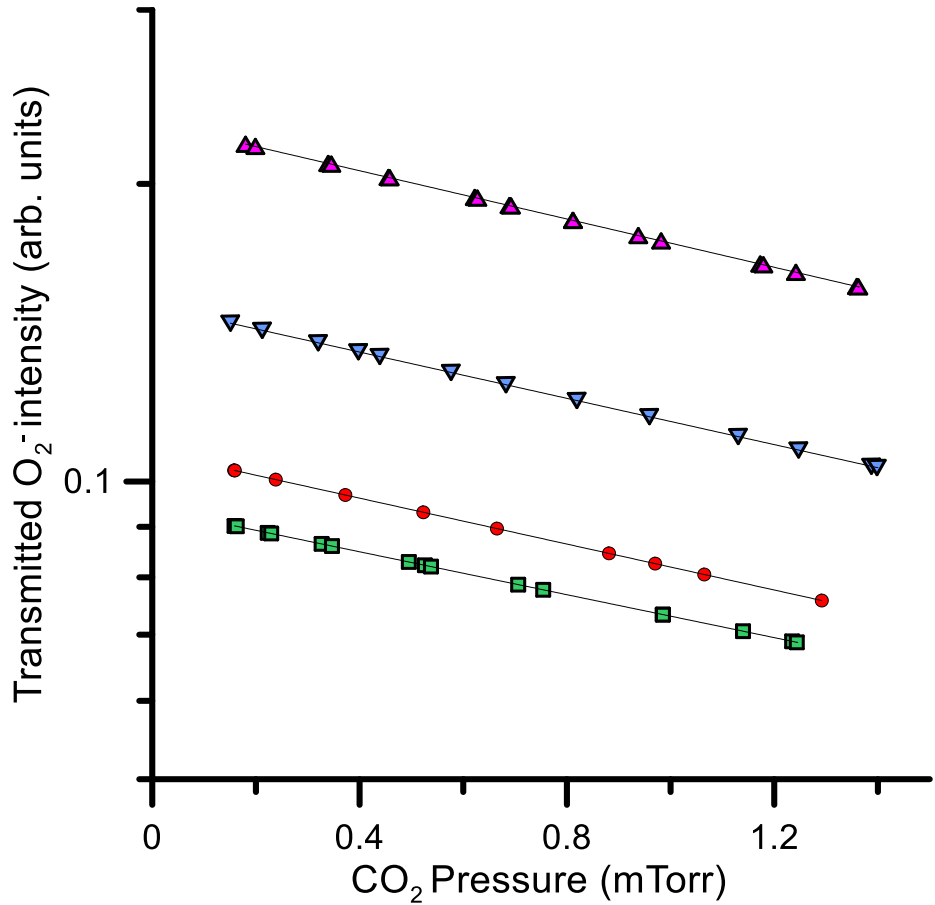

**FIG. 1.** Transmitted $O_2^-$ intensity as a function of the $CO_2$ pressure in the gas cell for 50 (●), 250 (■), 500 (▲) and 900 (▼) eV impact energies. Semilogarithmic fits of the experimental data are also plotted (—).

On the other hand, the TICS ($\sigma_+$) is derived from the ratio between the total positive ion detected intensity and the primary anion intensity, according to the expression:

$$\sigma_+ = \frac{I_+}{I_-} \frac{1}{nl} \quad , \tag{2}$$

where $I_+$ is the total positive ion intensity as recorded by the MCP1 detector, $I_-$ is the $O_2^-$ intensity measured by the MCP2 detector (transmitted $O_2^-$ intensity), $n$ is the molecular gas density derived from the gas pressure ($P$) and the temperature ($T$), assuming an ideal gas behavior, and $l$ the effective ion extraction length which is the diameter of the ion extraction aperture (2 mm). In these conditions, Eq. (2) provides absolute values of the integral ionization cross sections. Under the present experimental conditions, the counting efficiency of both MCP1 and MCP2 is about 100%. However, as we cannot ensure that the primary beam intensity detected by MCP2 corresponds to the actual intensity of this beam in the gas cell, we can only obtain relative values of the ionization cross sections as a function of the anion impact energy.

### b. Uncertainty analysis

Concerning the TEDCS, statistical uncertainties have been constrained below 5% by repeating at least 5 times the anion attenuation measurements and their corresponding data acquisition and analysis procedures for each impact energy considered in this study. The accuracy of the pressure determination is given by the uncertainty limit of the MKS-Baratron gauge (1%, according to the manufacturer data). However, as mentioned above, due to pressure gradients in the gas cell we considered the gas pressure as the average of those measured at both sides of the cell, so introducing an additional error contribution to the experimental cross sections of about 7%. Adding *in quadrature* all the known error sources and statistical uncertainties we determine a total uncertainty limit of 8-10% for the absolute electron detachment cross sections. With respect to the relative ionization cross-section, instabilities in the hollow cathode discharge led to statistical uncertainties of 7-14%, depending on the impact energy. Combining all the above error sources, the total uncertainty limit for these measurements has been determined to be within 10-16%.

### 3. Computational Details

We have thoroughly explored the potential energy surfaces of the neutral, anionic and cationic species involved in the collision, including different spin multiplicities. The geometry optimizations were performed using the Møller–Plesset perturbation theory, a post-Hartree–Fock ab initio method, including perturbations up to second order (MP2) [34] in combination with the aug-cc-pVTZ basis set [35]. We further computed the harmonic frequencies over the optimized geometries at the same level of theory to verify that they are minima in the corresponding potential energy surface and to obtain thermochemical values [36]. More accurate relative energies for each considered specie were obtained by means of Coupled Cluster calculations [37] including single and double excitations and the triplet ones perturbatively, CCSD(T), with the same basis set aug-cc-pVTZ, also over the geometry previously optimized. All these calculations were performed with the Gaussian16 program [38].

## 4. Results and Discussions

### 4.1. Induced fragmentation

Fig. 2 shows the TOF mass spectra for the cation fragmentation induced by collisions of $O_2^-$ with $CO_2$ at the impact energy of 600 eV. The assigned masses of the detected species are tabulated in Table I. The $CO_2$ cationic fragmentation has been analyzed for impact energies ranging from 150 to 900 eV by measuring the intensity of each TOF feature, corresponding to each identified fragment, with respect to that of the parent ion ($m/z = 44$ u).

**Table I.** TOF features formed in $O_2^-$ collisions with $CO_2$ and tentative cationic species assignation.

| Mass(m/z) | Proposed assignment |
|---|---|
| 16/17 | $O^+$ |
| 27/28 | $CO^+$ |
| 44 | $CO_2^+$ |
| 62-64 | $(CO_2 \cdot O)^+$ |
| 74-76 | $(CO_2 \cdot O_2)^+$ |

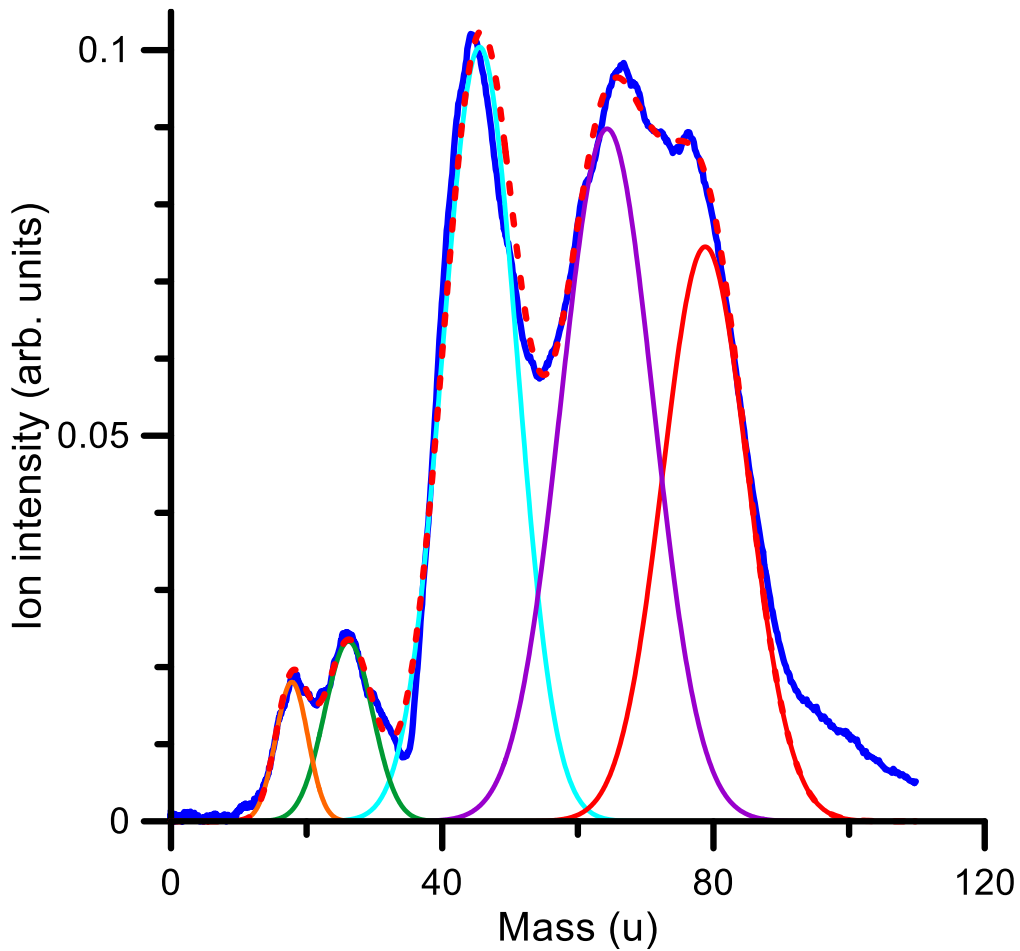

**FIG. 2.** TOF mass spectrum of the positive ion induced fragmentation of $CO_2$ in collisions with 650 eV oxygen anions ($O_2^-$). Gaussian fitting analysis: $m/z(u)$ = 16/17 (—); 27/28 (—); 44 (—); 60-64 (—); 73-77 (—). (See also the proposed cation assignments shown in Table I)

A deeper analysis of Fig. 2 reveals that the most intense cationic fragment at 650 eV can be assigned to the parent ion $CO_2^+$ and those fragments with 16/17 $m/z$ and 27/28 $m/z$ would correspond to $O^+$ and $CO^+$ fragments, respectively. Surprisingly, two structures with $m/z$ greater than the parent ion ($m/z$ around 60-64 and 74-76) have also been observed in the mass spectra. These fragments have been tentatively assigned to the formation of $(CO_2 \bullet O)^+$ and $(CO_2 \bullet O_2)^+$ complexes. At this point, it is interesting to compare the fragmentation induced by the anion beam to that produced by the electron beam (described in section 2) in similar collision conditions. The results are shown in Fig. 3. where both spectra are normalized to the parent ion ($m/z=44$) intensity. As this figure shows $O_2^-$ projectiles produce a higher m/z fragmentation than electrons. The main difference corresponds to the formation of molecular compounds with masses higher than that of the parent ion ($m/z>44$). The intensity enhancement of fragments with $m/z \cong 28$ suggests that these high-mass compounds could decay via $CO^+$ formation. Understanding the origin of these fragments ($m/z \geq 44$) represents a major challenge and requires a detailed theoretical analysis to recognize the possible processes and interactions behind their appearance (see section 4.2). Nevertheless, we have previously obtained similar results in our recent study on $O_2^-$ collisions with benzene [32], where the formation of large complex molecules has been reported. In that joint experimental and theoretical study [32], we proposed the formation of $C_6H_6O_2^+$ via a sudden double ionization of benzene and the subsequent electrostatic attraction between the $O_2^-$ and the formed dication, leading to stable diol structures. This mechanism is consistent with further partial and total ionization cross section measurements [33].

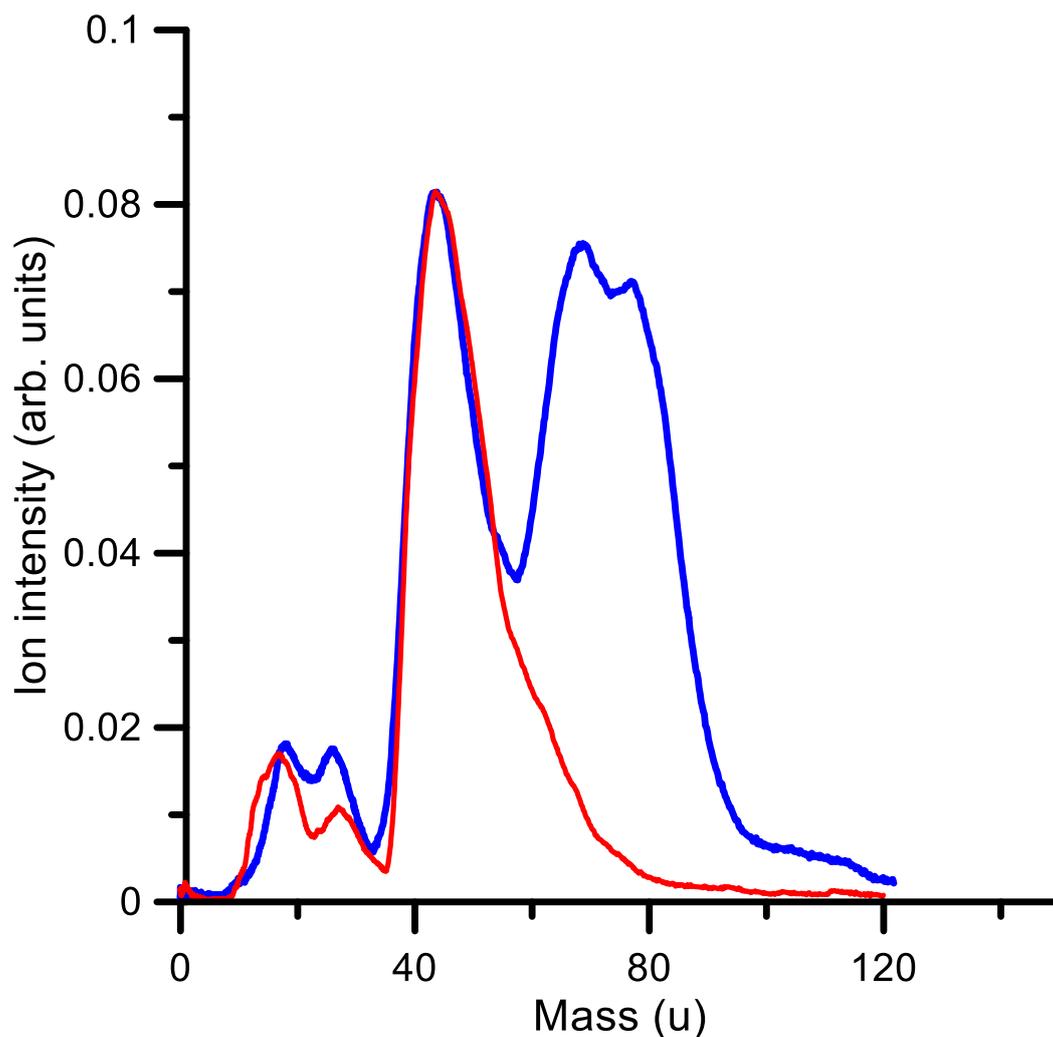

**FIG. 3.** Positive ion induced fragmentation by the superoxide anion beam (—) and the electron beam (—), respectively, both at 600 eV impact energy.

To analyze the variation of each fragment intensity as a function of the collision energy, Fig. 4 shows the observed relative intensity of the positive ion fragments produced by the collision of $O_2^-$ with $CO_2$ for the whole impact energy range considered in this study (160-900 eV). For impact energies below 600 eV, the most intense fragment is found to be the parent ion, followed by those fragments with higher masses (m/z > 44). However, between 600 and 900 eV, the intensity of those fragments increases and is comparable to that of the parent ion. Therefore, we can expect that the major contribution to the total ionization cross section, in the investigated energy range, results from the formation of the parent ion (*m/z*=44) and the high-mass compounds, especially in the 600-900eV impact energy range.

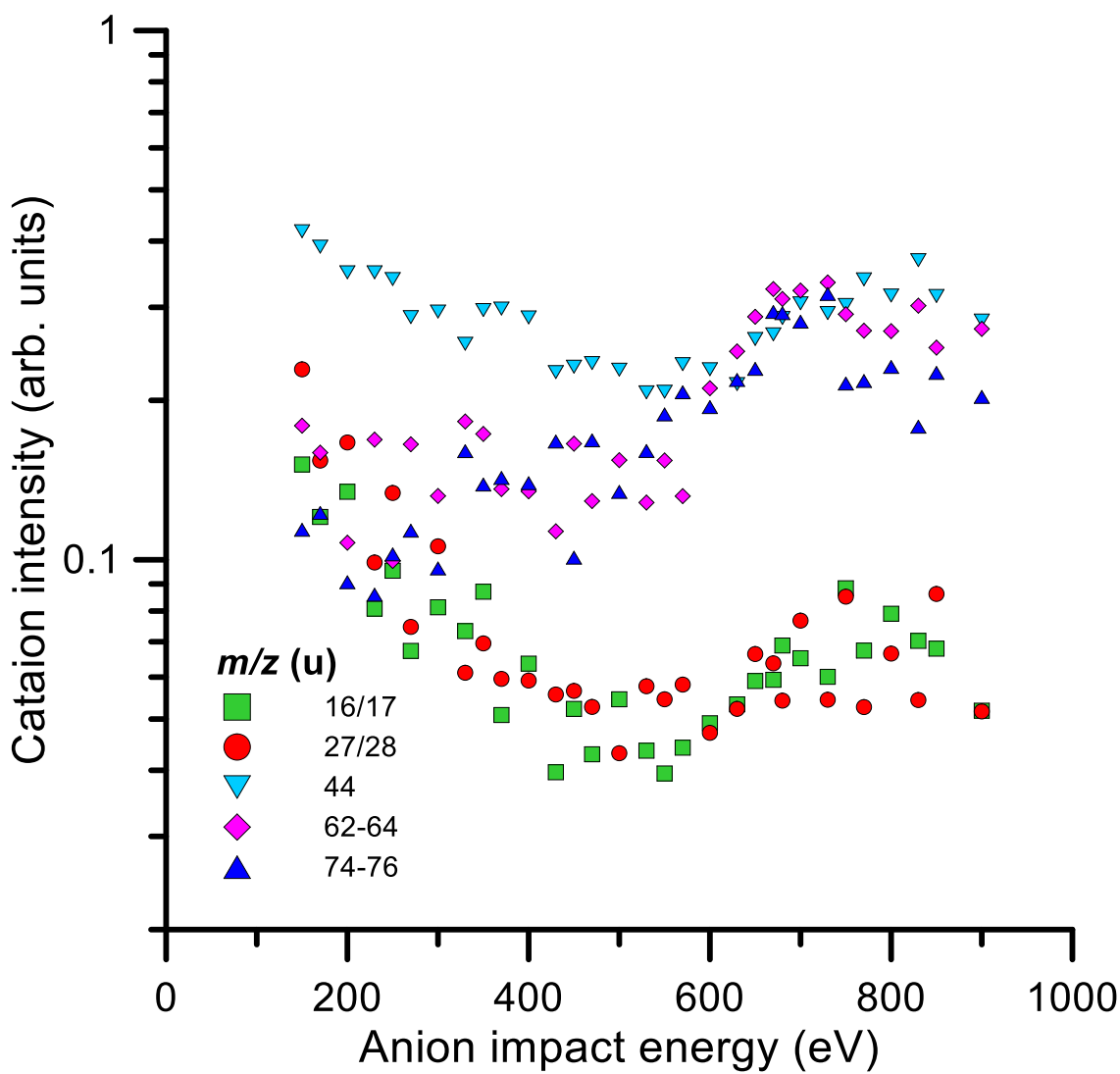

**FIG. 4.** Relative intensity of the different cation species detected by the TOF mass spectrometer: ■, $O^+$; ●, $CO^+$; ▼, $CO_2^+$; ◆, $(CO_2 \bullet O)^+$; ▲, $(CO_2 \bullet O_2)^+$

## 4.2. Structures and mechanistic proposal

Fig. 5 shows a mechanistic proposal for the formation of high mass cationic species. We present in the figure relative enthalpies, ΔH, for different points in the potential energy surfaces, refereed to the entrance channel $CO_2 + O_2^-$. Detection of cationic species with high mass implies ejection of two electrons accompanied with the reaction of the ionized formed species. Double ionization channels can be achieved by direct emission of the two electrons or in a sequential two-steps mechanism: single ionization in a high-energy state followed by autoionization leading to the doubly ionized channels. The highest energy channel that we have considered corresponds to a situation in which the two electrons are ejected from the $CO_2$, and it is found at almost 40 eV with respect to the entrance channel. Charge exchange processes would lead to $CO_2^+$, detected in the experiment ($m/z$=44). We also identify oxygen exchange channels, where $CO^+$ ($m/z$=28) and $CO_3^+$ ($m/z$=60) appear at c.a. 19 and 21 eV respectively. Finally, we also show the relative energy range in which formation of the complex $[CO_2 \bullet O_2]^+$ or $CO_4^+$ species takes place, c.a. 12 – 19.5 eV. The corresponding computed structures are given in Fig. 6. Some of these correspond to compounds bonded through van der Waals forces; the rest show covalent bonds stabilizing a $CO_4^+$ structure. Both kind of structures have been found in the potential energy surfaces with doublet and quartet spin multiplicity.

The experimental detection of $CO^+$, $CO_3^+$ and $CO_4^+$ implies population of channels well above the ground state of the doubly ionized channels, $CO_2(^1\Pi_u) + O_2^+(^2\Pi_g)$. Therefore, the energy appearance of such species in the calculations indicates that highly charged $CO_2^{2+}$ and/or highly energetic states of the $[CO_2 \cdot O_2]^+$ complex are populated in the collision following double ionization. One possible path to form the experimentally detected complexes would imply the electrostatic attraction of $CO_2^{2+}$ and $O_2^-$ forming the $[CO_2 \bullet O_2]^+$ complex. Other possible path could be the evolution of states with a high internal energy into $CO_4^+$ stable structures. In both cases, $CO_4^+$ and $CO_3^+$ channels can be energetically achieved.

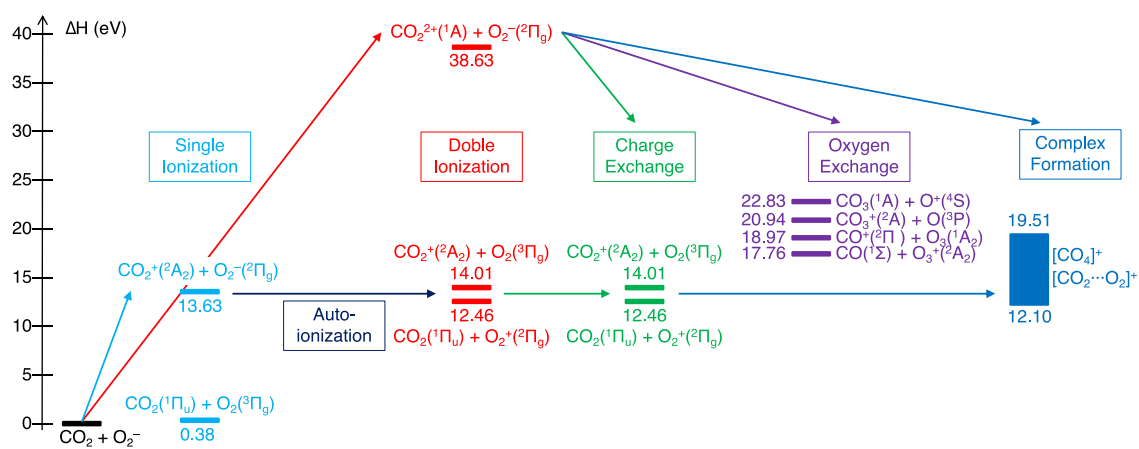

**FIG. 5.** Relative energies (enthalpy) of the different species involved in the mechanisms referred to the initial reactants in the collision $CO_2 + O_2^-$. Different colors denote different processes or final products.

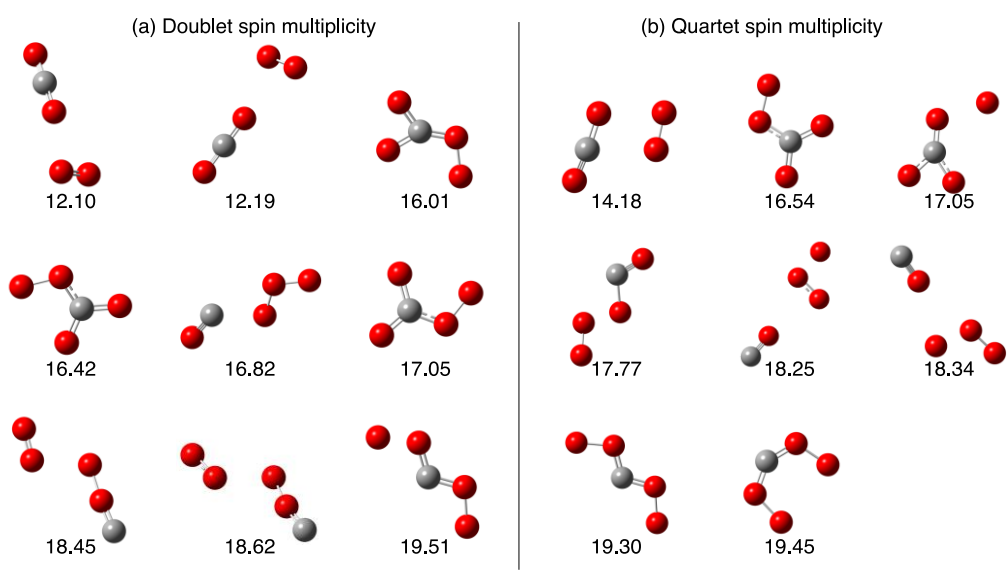

**FIG 6.** Optimized structures of the exit channels for the [CO$_4$]$^+$ molecules (or [CO$_x$·O$_{4-x}$]$^+$ complexes), with the two possible spin multiplicities. Numbers show the relative energy (enthalpy in eV) of the corresponding structure referred to the initial reactants in the collision CO$_2$ + O$_2^-$.

### 4.3. Total electron detachment and total ionization cross sections

The absolute total electron detachment cross sections (TEDCS) in the energy range 50-950 eV, together with the relative total ionization cross section (TICS) for energies ranging from 160-900 eV, obtained with the experimental procedures described above, are shown in Fig. 7. Numerical data are also presented in Table II and Table III, respectively. A close inspection on this figure shows that the TEDCSs sightly decrease in magnitude while energy increases up to 250 eV and subsequently increases up to 550 eV, reaching a kind of plateau for higher energies. This behavior is consistent with previous results for collisions of different positive and negative ions with CO$_2$ [17, 19, 22, 39]. In addition, it is observed a feature like-a-shoulder between 300 and 450 eV, with a local maximum placed at around 400 eV. Although the origin of this feature it is not clear, it could be in part associated with ionization processes as we will discuss below. On the other hand, in Refs. 20, 21 the role of charge exchange in the formation of resonances was studied by assuming the collision system as a wave full of free electrons affecting the molecule, showing that this resonance formation would occur at relatively high energies, at around 2 keV and 100 eV in the cases of CO$_2$, and O$_2$, respectively. Similarly, Huq [19] obtained results in full agreement with Tuan et al. [20] and they concluded that multiple resonances are present in the reaction H$^-$ + CO$_2$, noting the remarkable and evident influence of $^2\Pi_u$ resonance states on electron detachment dynamics for collision energies above 100 eV. This would be reflected in the increase of the TEDCS that we observe in the present study, thus revealing that the direct detachment of electrons would be the predominant process at lower collision energies.

**Table II.** Present experimental results of the TEDCS for O$_2^-$ anion collisions with CO$_2$ together with their associated uncertainties.

| Energy (eV) | TEDCS (×10$^{-20}$ m$^2$) | Total uncertainty (±) |
|---|---|---|
| 50 | 23.53 | 1.57 |
| 70 | 23.38 | 1.66 |
| 100 | 23.04 | 1.55 |
| 130 | 23.03 | 1.54 |
| 150 | 23.01 | 1.79 |
| 180 | 22.80 | 2.47 |
| 200 | 22.68 | 1.86 |
| 230 | 22.59 | 1.03 |
| 250 | 22.25 | 2.09 |
| 270 | 22.46 | 1.08 |
| 300 | 22.72 | 2.33 |
| 330 | 23.13 | 1.52 |
| 350 | 23.39 | 1.08 |
| 370 | 23.59 | 1.20 |
| 400 | 23.81 | 1.71 |
| 430 | 23.77 | 1.49 |
| 450 | 23.71 | 1.89 |
| 470 | 24.00 | 1.69 |
| 500 | 24.62 | 2.20 |
| 550 | 24.96 | 1.11 |
| 600 | 24.80 | 2.42 |
| 650 | 24.72 | 1.85 |
| 700 | 24.57 | 1.96 |
| 750 | 24.48 | 2.00 |
| 800 | 24.54 | 2.25 |
| 850 | 24.46 | 1.97 |
| 900 | 24.41 | 2.26 |
| 950 | 24.27 | 1.76 |

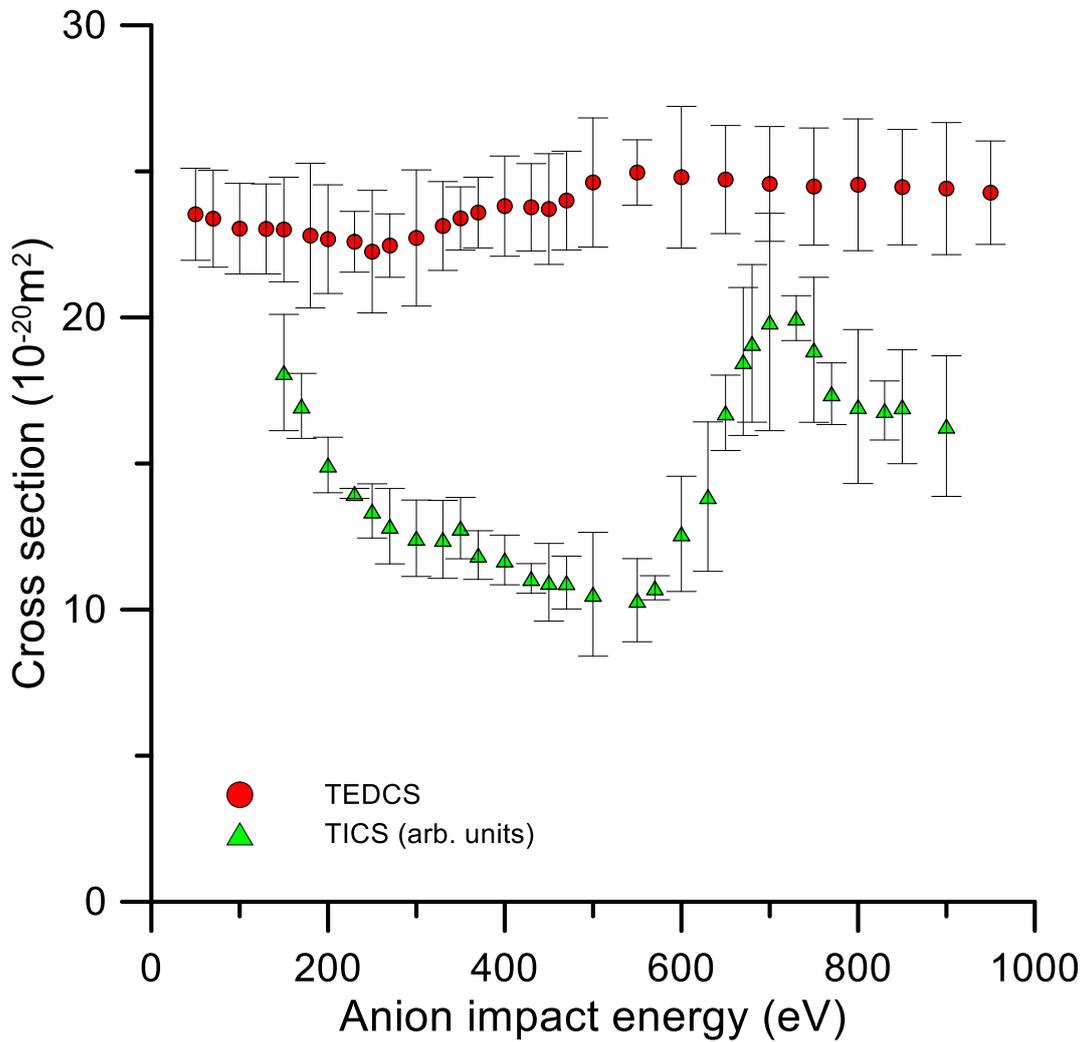

**FIG. 7.** TEDCS (●) in the 50–950 eV energy range for $O_2^-$ collision with $CO_2$, and relative TICS (◆) for the impact energy range 150 y 900 eV for the formation of positives ions. Note that the TICSs are just relative values in arbitrary units.

In contrast, the results we obtained for the relative TICS from the measured positive ion yields (see Table II) show a different behavior. They sharply decrease for increasing energies up to 550 eV, then notably increase up to 700 eV and again decrease for higher energies. The TICS profile exhibits two local maxima at around 350 eV and 730 eV (see

Fig. 4). The local maximum found in the TICS at 350 eV close to the shoulder in the TEDCS above 300 eV we can consider both as being partially associated. Similar behavior has been notd in our previous study on $O_2^-$ collisions with $N_2$ [16]. The second local maximum in the TICS is found around 700 eV. Since we only obtained relative values it is difficult to evaluate its correlation with the corresponding TEDCS. However, as derived from the partial ICS shown in Fig. 5, the main contributions to this maximum come from the formation of the higher mass complexes assigned to $(CO_2 \bullet O)^+$ and $(CO_2 \bullet O_2)^+$. We led to a similar conclusion in the case of $O_2^-$ collisions with benzene [33].

**Table III.** Present relative values of the total ionization cross section (TICS) for $O_2^-$ anion collisions with $CO_2$, together with their associated uncertainties.

| Energy (eV) | TICS (Arb. units) | Total Uncertainty (±) |
|---|---|---|
| 150 | 10.99 | 1.99 |
| 170 | 6.56 | 1.11 |
| 200 | 6.35 | 0.94 |
| 230 | 1.22 | 0.17 |
| 250 | 6.93 | 0.92 |
| 270 | 10.04 | 1.29 |
| 300 | 10.51 | 1.30 |
| 330 | 10.78 | 1.33 |
| 350 | 8.20 | 1.04 |
| 370 | 7.03 | 0.83 |
| 400 | 7.28 | 0.85 |
| 430 | 4.59 | 0.50 |
| 450 | 12.15 | 1.32 |
| 470 | 8.27 | 0.90 |
| 500 | 20.15 | 2.12 |
| 550 | 13.77 | 1.42 |
| 570 | 3.84 | 0.41 |
| 600 | 15.66 | 1.97 |
| 630 | 18.44 | 2.55 |
| 650 | 7.70 | 1.28 |
| 670 | 13.69 | 2.53 |
| 680 | 14.12 | 2.69 |
| 700 | 18.75 | 3.72 |
| 730 | 3.84 | 0.76 |
| 750 | 13.15 | 2.48 |
| 770 | 10.99 | 1.05 |
| 800 | 6.56 | 2.63 |
| 830 | 6.35 | 1.01 |
| 850 | 1.22 | 1.94 |
| 900 | 6.93 | 2.40 |

## 5. Conclusions

We have measured for the first time absolute TEDCS for $O_2^-$ collisions with $CO_2$ for impact energies ranging from 50 to 950 eV by using an anion beam transmission technique. Relative partial and total ionization cross sections in the energy range 150-900 eV have been determined by analyzing the mass of the cations formed during the collision processes by means of a TOF mass spectrometer. For impact energies up to about 500 eV the most intense cation signal corresponded to the parent ion ($CO_2^+$) formation. However, within the energy range (500-900 eV) intense ion signals corresponding to higher mass species were dominant, increasing the TICS values to reach a local maximum around 700 eV. The mechanisms of formation of such species have been inferred from quantum chemistry calculations exploring different potential energy surfaces: double ionization is followed by electrostatic attraction forming $[CO_2 \bullet O_2]^+$ and $CO_4^+$ structures, as well as $CO_3^+$.

## Acknowledgements

The authors acknowledge the generous allocation of computer time at the Centro de Computación Científica at the Universidad Autónoma de Madrid (CCC-UAM). This work was partially supported by MICINN (Spanish Ministry of Science and Innovation) projects PID2019-104727RB-C21 and PID2022-138470NB-I00 funded by MCIN/AEI/10.13039/501100011033, and the "María de Maeztu" (CEX2023-001316-M) Program for Centers of Excellence in RD. This study has been performed within the framework of the 21GRD02 BIOSPHERE project supported by the European Association of National Metrology Institutes (EURAMET).